\documentclass[referee, sn-mathphys-num]{sn-jnl}

\usepackage{graphicx}%
\usepackage{multirow}%
\usepackage{amsmath,amssymb,amsfonts}%
\usepackage{amsthm}%
\usepackage{mathrsfs}%
\usepackage[title]{appendix}%
\usepackage{xcolor}%
\usepackage{textcomp}%
\usepackage{manyfoot}%
\usepackage{booktabs}%
\usepackage{algorithm}%
\usepackage{algorithmicx}%
\usepackage{algpseudocode}%
\usepackage{listings}%
\usepackage{subfig}
\usepackage{graphbox}
\usepackage{enumitem}
\usepackage{wrapfig}
\usepackage{lineno}

\begin{document}
\title[Article Title]{Improving the Fairness of Deep-Learning, Short-term Crime Prediction with Under-reporting-aware Models}



\author[2]{\fnm{Jiahui} \sur{Wu}}

\author[1]{\fnm{Vanessa} \sur{Frias-Martinez}}

\affil[1]{
\orgname{University of Maryland}}

\affil[2]{ \orgname{Huawei}}



\abstract{Deep learning crime predictive tools use past crime data and additional behavioral datasets to forecast future crimes. Nevertheless, these tools have been shown to suffer from unfair predictions across minority racial and ethnic groups.  
Current approaches to address this unfairness generally propose either pre-processing methods that mitigate the bias in the training datasets by applying corrections to crime counts based on domain knowledge or in-processing methods that are implemented as fairness regularizers to optimize for both accuracy and fairness. 
In this paper, we propose a novel deep learning architecture that
combines the power of these two approaches to increase prediction fairness. 
Our results show that the proposed model improves the fairness of crime predictions when compared to models with in-processing de-biasing approaches and with models without any type of bias correction, albeit at the cost of reducing accuracy.}

\keywords{}



\maketitle

\section{Introduction}
\label{s3-sec:intro}

Crimes negatively impact the wellbeing of individuals and society as a whole. In 2020, the US saw a significant crime rise across major cities\footnote{https://www.cnn.com/2021/04/03/us/us-crime-rate-rise-2020/index.html}.
Researchers from various fields such as criminology, geographic information science, urban planning and data science, have conducted studies about the patterns of urban crimes. These studies help us better understand when and why certain crimes might happen and, more importantly, provide insights into the design of interventions to reduce the volumes of crimes.
One critical research direction of such efforts is place-based crime prediction that focuses on predicting the number of crime incidents or crime occurrence for a given location.
Environmental criminology provides theoretical foundations to study crimes from the perspective of places \cite{Evans2002-xz, Weisburd2012-lm}. Places with different urban functions can be viewed as crime attractors and crime generators \cite{Brantingham1995-ra}. 
Through the lens of place-based crime prediction, we can study the complex relationship between future crimes and historical crimes, built environment and social interactions in different places. 

Place-based crime predictions are typically carried out using either long-term or short-term approaches. 
Long-term crime prediction analysis, such as monthly or annual crime prediction, allows us to understand how the environmental factors of places shape future crimes; and in turn, help us inform better urban planning that improves the urban environment potentially decreasing crime occurrence. On the other hand, short-term crime prediction analysis focuses on next-day crime prediction {\it i.e.,} the identification of places where there will be crimes the next day. Short-term crime prediction is generally used to better allocate policing resources to response to crimes more swiftly. 
In this study, we focus on short-term crime prediction analysis.

Various models have been developed to tackle this problem. From kernel density estimation  - which was very common in the early efforts of crime prediction
\cite{Chainey2008-ys}, to  
epidemiological models 
whereby the spatio-temporal patterns of crimes in one location increase the probability of other incidents occurring at nearby locations \cite{Johnson2014-fc, Mohler2015-gm}.
Nevertheless, more recent deep learning approaches have shown superior performance in short-term crime prediction by modeling the spatio-temporal patterns of crime in the built environment as non-linear patterns \cite{Duan2017-fg, Huang2019-sx, Wu2020-ml}.
Based on the \textit{Crime opportunity theory} that suggests that human mobility is a key factor in crime generation \textit{i.e.}, the higher the presence of people or property, the more crimes could happen, recent work has also shown that deep learning architectures enhanced with mobility data characterizing local mobility patterns can improve the accuracy of the predictions \cite{wu2022enhancing}.


With the increasing application of predictive modeling in high stakes social impact settings, algorithmic fairness has become a critical component of predictive systems. Algorithmic fairness, which focuses on understanding and correcting bias in data and algorithms, is especially important for short-term crime prediction models as these models might influence the allocation of public resources such as police patrol scheduling. 
The debate over data bias issues in crime incident datasets is almost as old as the crime datasets themselves \cite{Levitt1998-vl}. 
Quantitative work has shown that bias might be present in crime data due to under-reporting and under-recording issues. For example, low-income and female-headed households are related to crime under-reporting \cite{Varano2009-rl}; and research has revealed police under-recording of crimes associated to certain demographics \cite{Wu2020-ty}. 
As a result, crime predictive algorithms have been shown to replicate and sometimes exacerbate the bias present in the training crime data during the prediction stage. 
For example, 
recent work has shown that unfair predictions are pervasive across minority racial and ethnic groups when deep-learning, short-term crime prediction models are used, and that these unfair predictions can be explained by the inherent data bias in the crime incident data \cite{wu2023auditing}.


Approaches to address unfair place-based predictions in deep learning models, for crime contexts and beyond, generally propose
pre-processing or in-processing approaches. Pre-processing methods attempt to mitigate the bias in the training datasets by  applying correction methods to crime counts, before the data is fed into the model for training\cite{jiang2024graph, jagodnik2020correcting}. The correction methods take advantage of domain knowledge of determinants for under-reporting  ({\it e.g.,} poverty, unemployment rate) to infer the actual \textit{true} crimes. 
In-processing methods, on the other hand, are generally implemented
as fairness regularizers that modify the loss function to optimize for both accuracy and fairness in the predictions during 
 model training \cite{Yan2019-iz}. As a result, in-processing methods do not attempt to correct the crime statistics to its true value, but rather use the crime datasets "as-is" and correct the bias embedded in the predictions. 
In this paper, we propose a novel under-reporting-aware deep learning method for short-term crime prediction that combines the power of correction methods and optimization methods to increase prediction fairness. 
Our predictive model attempts to model and correct the under-reporting processes that affect crime datasets while training the prediction model to accurately predict the true crime statistics. Specifically, 
we propose a convolutional gate mechanism to model the crime (under-)reporting process and infer actual \textit{true} crime reporting rates using under-reporting determinants. These true crime reporting rates are then used to modify the crime predictions of a deep learning model during training.

We evaluate the fairness and accuracy of the proposed approach, and compare it against under-reporting-unaware baselines. 
We use publicly available fine-grained crime and human mobility data based on a large-scale mobile phone dataset from the US \cite{kang2020multiscale}. The experimental evaluation is done across four American cities (Austin, Baltimore, Chicago and Minneapolis) and for multiple types of crimes, because crime patterns might differ across geographies and types of crimes. 
Our main contributions are:

 \begin{itemize}
    \item A novel under-reporting-aware deep learning method for short-term crime prediction that attempts to model and correct the under-reporting processes that affect crime datasets while training the prediction model to accurately predict the true crime statistics. Our results show that the proposed model improves the fairness of crime predictions 
    when compared to models with in-processing de-biasing approaches and with models without any type of bias correction, albeit at the cost of reducing accuracy.
    \item A high-quality experimental evaluation by looking into fairness and prediction accuracy for four cities in the US with diverse demographic characteristics: Baltimore, Minneapolis, Austin and Chicago, and for different types of crime. 
    
 \end{itemize}

\section{Related Work}

\subsection{Crime Prediction with Mobility Patterns}
Historical crime data and socioeconomic data are often used in crime prediction models \cite{Catlett2019-ra}.
For example, historical crime hotspots can be used to assess the risk of future crimes 
\cite{Mohler2015-gm, Catlett2019-ra}.
Mohler uses a marked point process to model the dependency between gun crimes and homicides for homicide prediction in cities \cite{Mohler2015-gm}. 
Neural networks have also been utilized to model the spatio-temporal patterns in historical crimes for future crime prediction \cite{Wu2020-ml}.
In addition to historical crimes, census data \cite{Kadar2018-ku} and points of interest (POI) \cite{Zhao2017-da} have also been used to enhance crime prediction. 
The proliferation of human mobility data, such as mobile phone data, geo-located social media, taxi pick-up/drop-off and check-ins, has allowed for the use of mobility features to predict crime incidents. One of the most common mobility feature used in crime prediction is \textit{footfall} defined as the number of individuals present in a given area at a given time span. 
Various studies use footfall as a feature to predict future crimes \cite{Bogomolov2015-pd, Kadar2018-ku}.
Bogomolov {\it et al.} estimate footfall and population diversity such as gender and age from mobile phone data and predict whether a regular grid cell will have a high or low level of crimes in the following month \cite{Bogomolov2015-pd}; while 
Kadar and Pletikosa extracted footfall from check-ins, subway and taxi data, along with other census and POI features, to predict the number of crimes for a given census tract using tree-based machine learning models \cite{Kadar2018-ku}.
Another mobility feature used in crime prediction contexts is the origin-destination matrix (OD) that characterizes human mobility (flows) between census tracts. Human mobility data has been used to characterize human behaviors in the built environment 
~\cite{vieira2010querying,hernandez2017estimating,frias2013cell,rubio2010human,wuspatial}, 
for public safety~\cite{wu2022enhancing,wu2023auditing}, during epidemics and 
disasters~\cite{wesolowski2012quantifying,bengtsson2015using,hong2017understanding,isaacman2018modeling,ghurye2016framework,hong2020modeling, Erfani_Frias-Martinez_2023,abrar2023analysis}, as well as to support decision making for socio-economic development
~\cite{frias2010socio,fu2018identifying,frias2012mobilizing, hong2016topic,frias2012computing,hong2019characterization}.
In this paper, we will focus on deep learning crime prediction models that exploit the predictive power past crime data and OD mobility matrices \cite{wu2022enhancing}.

\subsection{Under-reporting in Crimes Statistics}
Concerns about under-reporting in crime data are highly related to the production of the reports themselves. Although crime reporting systems around the world vary a lot, in a simplified way, we can identify two main phases: a crime first needs to be reported to the police by an individual, and it then needs to be recorded as a crime entry into the police database. 
When crimes are reported, around 80\% of them are reported by victims or witnesses, while the police on scene reports about 6\% and the rest are reported by offenders, alarm systems or officials other than police, among others \cite{Davidson1981-ib, Hart2003-ku}. 
However, there are various reasons why the public might choose not to report a crime. The crime being "too trivial/no loss" used to be the most important reason, but recently "Police could do nothing" has come on top \cite{Tarling2010-qw}. 
After an incident is reported, the police decides whether or not to record the incident as a crime event in the database.
Various factors can influence the police' decision such as insufficient evidence and/or individual biases \cite{Mason2019-ky}
As a result, under-reporting in crime is heavily impacted by social disparities. For example, in Kensington, middle-class crime complaints are more likely to be reported and accepted by the police (\textit{i.e.,} high reporting rate and high recorded rate), while the reports from white working-class tend to be rejected (low recorded rate) and racially-mixed communities are less willing to report (low reporting rate) \cite{Davidson1981-ib}.

Therefore, it is critical to address the existing bias in reported crime data so that crime predictions are fair across social groups. 
Although the reporting and recording of crime incidents are two different phases, in this study, we make no distinction between them as it is almost impossible to obtain such information from local police force. Instead, we simplify and quantify the under-reporting issue of crimes as the reporting rate, which is the ratio of the number of reported crimes in the police database to the number of (unobserved) true crimes that have occurred. This simplification is common in the literature 
\cite{Hart2003-ku}.


\subsection{Algorithmic Fairness}
There exists a plethora of computational algorithms making decisions with high societal impact such as loan requests, crime prediction or criminal sentencing. As a result, algorithmic fairness or the design of algorithms that treat social groups similarly, becomes a critical component of any predictive approach. 
Algorithmic fairness, especially the most commonly used notion of group or statistical fairness, is based on the notion of protected or sensitive attributes, such as gender and race (minority and non-minority). 
A protected attribute usually represents a population sub-group that has historically suffered from discrimination and therefore some form of (approximate) parity or non-discrimination regulation in the predictive algorithm is desired for these protected groups 
\cite{Chouldechova2018-jv}.
Fairness is a complex concept and there are different and sometimes conflicting definitions and thus a variety of fairness metrics \cite{Verma2018-kq}. Although the definitions of fairness vary, it has been empirically shown that there is usually a trade-off between the accuracy and fairness of prediction, {\it i.e.,} improvement in fairness is generally at the expense of the algorithmic accuracy 
\cite{Berk2017-nu}. 
In this paper, we propose a novel deep learning architecture to correct under-reporting in crime data while controlling for fairness across protected attributes and accuracy. By properly incorporating domain knowledge about potential under-reporting  - which is a source of data bias - we will show that we can improve fairness the fairness of crime prediction algorithms.

\section{Data}
\label{s3-sec:Dataset}
In this paper, three types of data are used: census data from the American Community Survey (ACS)\cite{us_acs}, crime incidents and human mobility. Next, we describe these and provide general statistics for the four cities evaluated in this study: Baltimore (Bal), Minneapolis (Min), Austin (Aus) and Chicago (Chi). These four cities were chosen based on the diversity of their demographics, as shown in Table \ref{s3-tab:race}, with Baltimore having majority Black and African-American population, Minneapolis majority White, Austin has a high White and Latino and Hispanic population and Chicago with a balanced mix of White, Black and African-American and Hispanic and Latino communities. Replicating the short-term crime prediction and fairness analysis across these four cities will provide a robust analysis across geographies. 

\subsection{ACS data}
The novel under-reporting-aware deep learning method for short-term crime prediction that we propose leverages domain knowledge of determinants for under-reporting to infer the actual \textit{true} crimes. 
The under-reporting determinants on the census-tract-level are socio-demographic variables obtained from 2019 American Community Survey (ACS) 5-year estimates \footnote{https://data.census.gov/}. 
The determinants we use are: poverty rate (PR), unemployment rate (UR), adult rate (AR), the percentage of people who are never married (never married rate, NMR), male to female ratio (M/F), percentage of female householder with children under 18 years old (FHHR), percentage of people who cannot speak English (linguistic isolation rate, LIR) and the percentage of foreign born population (foreign-born rate, FR). Additional details about these features are provided in Section 4. 

\begin{table}[]
\small
\centering
\begin{tabular}{cp{1.5in}p{1.5in}p{0.8in}p{0.8in}}
\toprule
    & \%   Not Hispanic or Latino, White Alone& \%   Black or African-American & \%   Hispanic or Latino & \% Asian \\
\midrule
Bal & 27.54\%                           & 62.46\%                & 5.12\%                   & 2.59\%   \\
Min & 59.80\%                           & 19.36\%                & 9.58\%                   & 6.13\%   \\
Aus & 49.08\%                           & 7.60\%                 & 33.64\%                  & 7.34\%   \\
Chi & 33.61\%                           & 29.48\%                & 28.89\%                  & 6.40\%  \\
\bottomrule
\end{tabular}
\caption{The percentage of population across race and ethnicity for the four cities according to the American Community Survey (2019 ACS 5-year estimates)\cite{us_acs}. The cities are: Baltimore (Bal), Minneapolis (Min), Austin (Aus) and Chicago (Chi).}
\label{s3-tab:race}
\end{table}

\begin{table}[]
\small
\centering
\begin{tabular}{@{}llllllll@{}}
\toprule
         &     & Jan    & Feb    & Mar    & Apr    & May    & Jun       \\ 
\midrule
         & Bal & 28.0\% & 27.2\% & 24.6\% & 22.4\% & 23.6\% & 25.0\% \\
Property & Min & 35.0\% & 33.4\% & 34.1\% & 35.3\% & 37.6\% & 34.7\% \\
Crime    & Aus & 32.9\% & 31.9\% & 30.6\% & 30.5\% & 31.2\% & 31.5\% \\
         & Chi & 23.5\% & 22.6\% & 19.7\% & 16.6\% & 19.6\% & 20.4\%  \\
\midrule
         & Bal & 21.6\% & 21.1\% & 21.8\% & 17.0\% & 21.6\% & 23.4\%  \\
Violent  & Min & 9.4\%  & 9.3\%  & 10.7\% & 8.5\%  & 10.3\% & 13.0\%   \\
Crime    & Aus & 4.0\%  & 3.7\%  & 4.5\%  & 4.2\%  & 5.0\%  & 5.4\%    \\
         & Chi & 11.5\% & 11.0\% & 9.9\%  & 8.3\%  & 10.2\% & 11.6\%  \\ 
\bottomrule
\end{tabular}
\caption{Crime occurrence monthly density for the four cities in 2020: Baltimore (Bal), Minneapolis (Min), Austin (Aus) and Chicago (Chi) in January through June.}
\label{s3-tab:crm_density}
\end{table}

\subsection{Crime incident data} 
The crime incident datasets for the four cities are obtained from their open data portals, covering crimes from January to December, 2020\footnote{ 
Bal: https://data.baltimorecity.gov/; 
Min: https://opendata.minneapolismn.gov/; \newline
Aus: https://data.austintexas.gov/; 
Chi: https://data.cityofchicago.org/;
}. 
Each crime incident is associated with the crime category it belongs to and with the time and location where it took place. Crime locations are generally geo-coded to the closest street or block in the city, however, to account for the potential spatial precision inaccuracy, 
We use a 50-meter buffer to associate crime incidents to urban census tracts (a similar approach has been implemented 
in prior work {\it e.g.,} \citet{Kadar2018-ku,De_Nadai2020-gt}).
Although crime incidents could be associated to smaller spatial units, the choice for spatial units is determined by the availability of human mobility data at the census tract level only. 
We group the crime incidents into two types: property and violent crimes, and we will evaluate short-term crime prediction and fairness for each type separately. Property crimes include arson, burglary, larceny-theft, and motor vehicle theft; while violent crimes include aggravated assault, forcible rape, murder, and robbery.
Tables \ref{s3-tab:crm_density} and \ref{s3-tab:crm_density2} show the monthly crime density for each city throughout 2020, where monthly crime density is computed as the percentage of census tracts with crime incidents during that month. 
The table shows that the four cities selected generally suffer from higher volumes of property crimes than violent crimes; and that they represent a diverse group with some cities suffering from higher volumes of violent and property crimes than others.

\begin{table}[]
\small
\centering
\begin{tabular}{@{}llllllll@{}}
\toprule
         &      & Jul    & Aug    & Sep    & Oct    & Nov    & Dec    \\ 
\midrule
         & Bal  & 24.0\% & 22.7\% & 24.7\% & 25.6\% & 24.2\% & 21.3\% \\
Property & Min  & 41.6\% & 43.3\% & 40.7\% & 41.3\% & 37.0\% & 33.2\% \\
Crime    & Aus  & 31.8\% & 34.3\% & 35.0\% & 33.3\% & 36.1\% & 34.4\% \\
         & Chi  & 22.5\% & 23.5\% & 22.2\% & 21.0\% & 19.7\% & 18.2\% \\
\midrule
         & Bal  & 23.2\% & 23.4\% & 22.4\% & 22.5\% & 21.1\% & 18.6\% \\
Violent  & Min  & 16.4\% & 14.6\% & 13.7\% & 12.9\% & 10.4\% & 8.3\%  \\
Crime    & Aus   & 5.7\%  & 5.3\%  & 5.2\%  & 4.7\%  & 5.3\%  & 5.2\%  \\
         & Chi  & 12.9\% & 12.8\% & 12.4\% & 11.1\% & 10.8\% & 9.3\%  \\ 
\bottomrule
\end{tabular}
\caption{Crime occurrence monthly density for the four cities in 2020: Baltimore (Bal), Minneapolis (Min), Austin (Aus) and Chicago (Chi) in July through December.}
\label{s3-tab:crm_density2}
\end{table}

\subsection{Human mobility data}
The pervasive presence of ubiquitous technologies such as smart phones, has allowed for the collection of large-scale human mobility data. Location intelligence companies like SafeGraph, collect pseudonymized mobile GPS location data using SDKs installed on individuals' mobile phones via mobile apps. SafeGraph offers multiple datasets. For this study, we use daily origin-to-destination flows at the census tract (CT) level from January to December, 2020. This dataset is publicly available (see \cite{kang2020multiscale}). To extract this dataset, SafeGraph assigns to each device a home location at the census block group level based on its night-time activity. Then, it tracks for each device all the trips from its home location to points-of-interest (POIs) in SafeGraphs' large POI database. Origin-destination (OD) flows are finally computed by transforming all the home-to-POIs trips to CT(O)-CT(D) trips and by computing the number of devices associated to each OD across all census tracts in a city. OD flow volumes are computed at a daily granularity. Since the devices in SafeGraph's database account for about 10\% of the entire population in the U.S., the OD flow volumes are re-scaled by the census population.

\begin{table}[]
\small
\centering
\begin{tabular}{p{1.8in}p{0.8in}p{0.8in}p{0.8in}p{0.8in}}
\toprule
 & Bal & Min & Aus & Chi \\
\midrule
Number of census tracts & 200 & 116 & 204 & 809 \\
\midrule
Volume of in-city OD flow & 4040.1 (1733.9) & 4004.3 (1653.7) & 8167.2 (3866.3) & 5307.3 (2821.6) \\
\midrule
Volume of out-of-city OD flow  & 1413.6 (1149.9) & 2055.8 (1749.5) & 2102.6 (1651.3) & 1198.9 (1646.3) \\
\midrule
The number of unique census tracts connected by in-city OD flow  & 38.7 (14.6)     & 30.5 (10.7)     & 66.8 (20.2)     & 61.0 (28.5)     \\
\midrule
The number of unique counties connected by out-of-city OD flow & 14.5 (11.9)       & 23.6 (20.8)       & 29.6 (17.2)       & 15.1 (20.5)      \\
\midrule
The number of unique states connected by out-of-city OD flow & 5.9 (3.3)       & 7.0 (4.2)       & 7.7 (4.0)       & 6.2 (4.0)      \\
\bottomrule
\end{tabular}
			
\caption{Human mobility flow statistics for the four cities under study: Baltimore (Bal), Minneapolis (Min), Austin (Aus) and Chicago (Chi). The numbers in each cell represent the mean (standard deviation) of the daily average across all census tracts in a given city in 2020. OD flows outside the city are flows that either start or end in a census tract that is not part of the city of interest.} 
\label{s3-tab:mobi_stats}
\end{table}

Table \ref{s3-tab:mobi_stats} shows general OD flow volume statistics for the four cities under study for the year 2020.
For each measure, the table shows the mean and standard deviation of its daily average values across all census tracts in each city.
In-city OD flows refer to flows whose origin and destination census tracts (CT(O) and CT(D)) are within the city; while out-of-city OD flows are flows in which either the origin or the destination census tract is outside the city under study.
To characterize mobility diversity, the table also shows the number of unique census tracts connected by in-city OD flows and the number of unique counties and states connected by out-of-city OD flows. 
It can be observed that most of the OD flows identified take place within the cities under study, with smaller volumes being associated to trips to counties outside the city, and even a smaller number to trips to other states. Consequently, there is a higher diversity in the number of distinct areas visited inside than outside the city (counties or states). 
A more detailed description of the features extracted from this dataset is covered in the next section.

\section{Under-reporting-aware Model to Improve Fairness of Short-term Crime Predictions}

\subsection{Problem setting}
\label{s3-sec:problem-setting}
In this paper, we focus on placed-based short-term crime prediction for urban areas. For that purpose, 
a city is divided into $N$ spatial units $\mathbf{S}=\{s_1, s_2, ..., s_N\}$ which for this study are defined as census tracts. Census tracts are chosen as spatial units because the human mobility flow dataset is only available at the census tract level. 
The short-term crime prediction is framed as determining whether there will be at least one crime the next day at a given census tract using prior crime and mobility data for that tract. 
Crime occurrences at a census tract $s_i$ on day $t$ are denoted as $h_{i,t}$ and $h_{i,t}=1$ is referred to as a crime hotspot. 

For each census tract $s_i$, two sets of daily predictive features are computed: 
1) historical crimes ($C$), defined as the daily number of past crime incidents; the input sequence for crime prediction at day $t$ is represented as
$\mathbf{C}_{i,t}=\{c_{i,t-T}, c_{i, t-T+1}, ..., c_{i, t-1}\}$ with  
$T$ being the length of the \textit{look-back} period {\it i.e.,} the time range used to characterize {\it history} and $c_{i, t-d}$ being the number of crime incidents $d$ days before day $t$;
and 
2) mobility features ($M$), defined as a set of ten daily features extracted from SafeGraph's daily OD matrices and denoted as
$\mathbf{M}_{i,t} = \{\mathbf{M}_{i,t}^{j} | j \in \{1, 2, ..., 10\}\}$ and $\mathbf{M}_{i,t}^{j} = \{m_{i,t-T}^{j}, m_{i, t-T+1}^{j}, ..., m_{i, t-1}^{j}\}$, where $m_{i,t-d}^{j}$ is the value of the $j$-th mobility feature at $d$ days before day $t$.
The ten features identified characterize mobility volumes and mobility diversity. 
Mobility volume features characterize the daily total number of people going in (inflow) and out (outflow) of a census tract within or outside the city under study, which have been shown to be related with the volumes of crime incidents \cite{Kadar2018-ku, Bogomolov2015-pd, Wu2020-ty};
while mobility diversity features characterize the regional influence, \textit{i.e.}, the number of unique regions visited by in/outflows, including census tracts, counties and states. Past research has shown that crimes committed by visitors are associated to different patterns (behaviors) than those of residents \cite{Boivin2018-cf}; and that pass-through traffic information improves crime prediction accuracy \cite{Kadar2020-ij}. Therefore, mobility diversity features are extracted to reflect the connections between the census tract $s_i$ and other regions.
Table \ref{s3-tab:ftr} shows a summary of all the features used in the short-term crime prediction models. Besides crime and human mobility data, I also add {\it Day of week} to the feature set to capture the difference between crime data and human mobility behaviors during weekdays and weekends.

\begin{table}[]
\centering
\small
\begin{tabular}{@{}ll@{}}
\toprule
Types     & Features\\
\midrule
Crimes    & Daily number of crimes\\
\midrule
Mobility  & Volumes of in-city inflow                   \\
Volumes   & Volumes of in-city outflow                   \\
          & Volumes of out-of-city inflow           \\
          & Volumes of out-of-city outflow             \\
\midrule
Mobility  & Number of CT connected by in-city inflow             \\
Diversity & Number of CT connected by in-city outflow            \\
          & Number of counties connected by out-of-city inflow  \\
          & Number of counties connected by out-of-city outflow \\
          & Number of states connected by out-of-city inflow    \\
          & Number of states connected by out-of-city outflow     \\
\midrule
day of week & Day of week \\
\bottomrule
\end{tabular}
\caption{Complete list of predictive (input) features for short-term crime prediction models. For census tract $s_i$, inflow (outflow) means $s_i$ is the destination (origin) of the OD flow.}
\label{s3-tab:ftr}
\end{table}


\textbf{Problem Statement}. Given the temporal sequences of input features ($C+M$) within the \textit{look-back} period $T$ for all census tracts in a city,
predict whether a census tract will be a hotspot (or not) in the next day $h_{i,t}=1, i\in[1,N]$ or $h_{i,t}=0$, otherwise.



\subsection{Proposed Architecture}

The crime reporting process can be described in two stages: 1) a crime incident is reported to the police and 2) the reported incident is recorded in the police database.
To model under-reporting that could happen in any of these two stages, we define the following variables. 
We define the number of \textit{true} crimes $y_{i,t}$ as the actual number of crimes that will occur regardless of whether they will be reported; and the reporting rate $\pi_i$ quantifies the under-reporting as the ratio of the number of reported crimes $z_{i,t}$ to the true crimes $y_{i,t}$, where $i$ refers to census tract $s_i$ and $t$ refers to day $t$.
In other words, $z_{i,t}=y_{i,t}\times pi_i$.
To model the crime-reporting process, $y_{i,t}$ is considered as a function of the predictive features for crimes, \textit{e.g.}, the features extracted from historical crimes and from the human mobility dataset; while the reporting rate, $\pi_i$, is defined as a function of the under-reporting determinants based on domain knowledge.

\begin{figure}
    \centering
    \includegraphics[width=\linewidth]{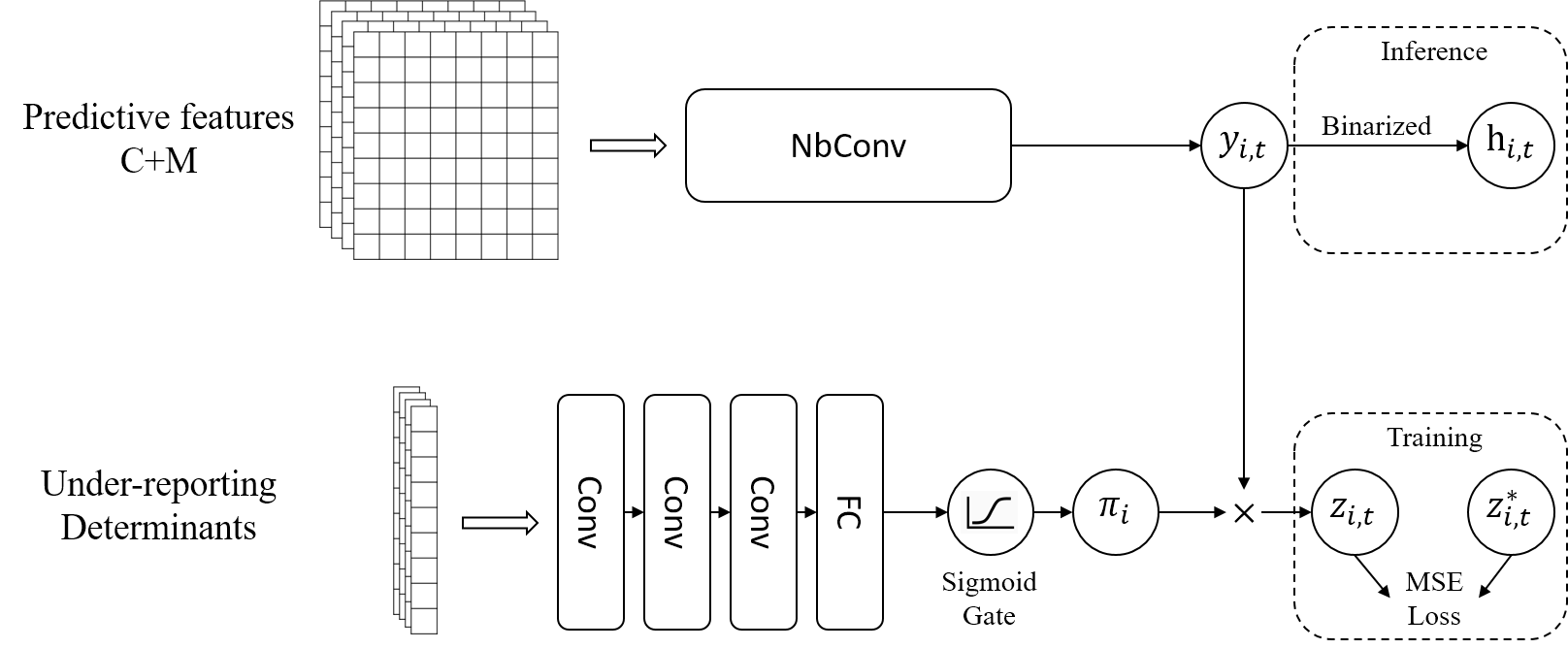}
    \caption{Under-reporting-aware short-term crime prediction with crime-reporting convolutional gate. The 2D feature maps for historical crimes, mobility features and under-reporting determinants are constructed based on the neighboring set for census tract $s_1$ in the same way as shown in Figure \ref{s3-fig:nbcnn_rearrange}.}
    \label{s3-fig:urgate}
\end{figure}

Our proposed under-reporting-aware short-term crime prediction model consists of two neural network branches, as shown in Figure \ref{s3-fig:urgate}.
The first (top) branch is the true crime predictor which infers the number of true crimes $y_{i,t}$ in the next day $t$ with predictive features for crimes \textit{i.e.,} past crimes and mobility features. We use a Neighbor Convolution (NBConv) model for this branch given its high performance for crime contexts\cite{wu2022enhancing}. 

\begin{figure}
    \centering
    \includegraphics[width=.45\linewidth]{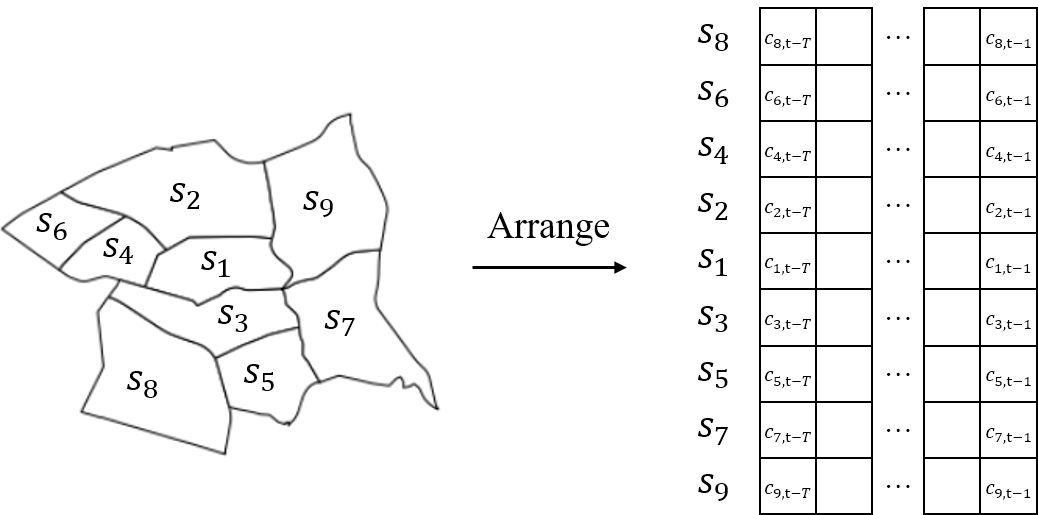}
    \caption{Arrange the nearest neighbors set for the target census tract $s_1$ and construct the 2D feature map for historical crimes. In the neighboring set of $s_1$, $s_2$ and $s_3$ is the closest to $s_1$; $s_4$ and $s_5$ are the next closest to $s_2$ and $s_3$ respectively; $s_6$ and $s_7$ are the next closest to $s_4$ and $s_5$; $s_8$ and $s_9$ are the next closest to $s_6$ and $s_7$. Similar process is applied to each of the ten mobility features.}
    \label{s3-fig:nbcnn_rearrange}
\end{figure}

Neighbor convolution models that account for spatio-temporal dependency have been used for crime prediction using historical data over a spatial grid \cite{Duan2017-fg}. To adapt this model to the  setting in this paper, where the spatial units are census tracts (non-regular division), we extract a fixed-length nearest neighbors set for each census tract for which the model outputs the next-day crime prediction. Specifically, we focus on the eight nearest census tracts for each target census tract. We arrange the target census tract in the middle and sort the nearest neighboring census tracts from closest to furthest to form a 2D feature map per input feature, as explained in Figure \ref{s3-fig:nbcnn_rearrange}. Such arrangement allows the kernel of the convolutional layer to model the spatio-temporal dependency through its local receptive field. These 2D feature maps are then input to the full convolution architecture. The original model in \cite{Duan2017-fg} contains inception and fractal blocks. In this paper, we discuss results for a model with only the first regular convolution blocks because it provided better performance than the full model.

The second (bottom) branch is a crime-reporting convolutional gate that is used to infer the reporting rate $\pi_i$ based on the under-reporting determinants, namely poverty rate (PR) and unemployment rate (UR) for property crimes; and poverty rate (PR), adult rate (AR), the percentage of people who are never married (never married rate, NMR), male to female ratio (M/F), percentage of female householder with children under 18 years old (FHHR), percentage of people who cannot speak English (linguistic isolation rate, LIR) and the percentage of foreign born population (foreign-born rate, FR) for violent crimes.
Since ACS provides not only the estimates but also the margin of error of the under-reporting determinants, we include both the estimates and margin of error in the feature maps for the convolutional gate to model the uncertainty in ACS estimates.
The convolutional gate proposed here not only models the non-linear relationship between the determinants and reporting rate $\pi_i$ of the target census tract $s_i$, but also between $\pi_i$ and the determinants of the neighboring set of $s_i$ to capture the spatial dependency.
The convolutional gate consists of three layers of convolutional blocks (Conv), a fully-connected layer (FC) and a Sigmoid gate.

\subsubsection{Training and Inference Process}
To obtain predictions for the binary variable $h_{i,t}$ (crime hotspot prediction), the under-reporting-aware model follows the training and inference process as shown in Figure \ref{s3-fig:urgate}.
First, the under-reporting-aware NbConv is trained in the regression setting with mean squared error (MSE) as the loss function 
$Loss = \frac{1}{N*TD} \sum_{i\in N} \sum_{t\in TD} \left( z_{i,t}^* - z_{i,t}\right)^2$, 
where $z_{i,t}^*$ is the ground truth and $z_{i,t}$ is the predicted number of reported crimes for census tract $s_i$ on day $t$, $N$ is the total number of census tract in a city and $TD$ is the number of days in the training period. 
In the training process, both the branch for true crimes $y_{i,t}$ and for reporting rate $\pi_i$ of the model are activated to infer the number of reported crimes $z_{i,t}$. Because only the ground truth of the reported crime incidents is available, the goal of the training process is to to minimize the error in predicting the number of reported crimes.
Then in the inference phase, 
only the top branch for $y_{i,t}$ is utilized to predict next-day crime hotspots $h_i$. $y_{i,t}$ is binarized as $h_{i,t}$ as follows:

\begin{equation}
    h_{i,t} = 
    \begin{cases}
    1, & y_{i,t} > \overline{y}_{t} \\
    0, & y_{i,t} \leq \overline{y}_{t}
    \end{cases}
    \label{s3-eq:yh}
\end{equation}

\begin{equation}
    \overline{y}_{t} = \frac{1}{N} \sum_{i\in N} y_{i,t}
\end{equation}

$\overline{y}_{t}$ represents the average predicted number of true crimes across all census tracts in the city on day $t$ and Equation \ref{s3-eq:yh} means that a census tract is considered a hotspot on the next day $t$ if the predicted number of true crimes is larger than the average of all census tracts.



\section{Evaluation Protocol}

\subsection{Experiment Setting}
Given the 1 year of data, we 
chronologically split the dataset into training (6.5 months), validation (0.5 month), and testing (5 month) sets. 
The validation set is used to tune the learning rate and early stopping {\it i.e.,} deciding the maximum number of epochs for training. Then, we re-train the model using the combination of training and validation set (a total of 7 months) and use the testing set to make next-day predictions (5 months).
The overall performance of a model is represented by its monthly F1 score, computed comparing the next-day crime prediction with the daily ground truth over all days for each testing month.
This experimental protocol with time series data has also been followed in other related work such as \citet{Huang2019-sx}.

\subsection{Baselines}
In order to evaluate the effects of modeling under-reporting on fairness and accuracy, we define three baselines. 
1) an "under-reporting-unaware" NbConv model using historical crimes and mobility features, denoted as UU.
2) an "under-reporting-unaware" NbConv model using historical crimes only, denoted as UU(C), to evaluate the potential impact of mobility features.
3) UU with the in-processing fairness enhancement method proposed in \cite{Yan2019-iz} which adds a fairness regularization to the loss function to minimize the per-capita score between protected and non-protected racial and ethnic groups. 
For this study, since the score corresponds to the predicted number of reported crimes, we compute the per-capita score as the predicted number of crimes divided by the population of the (non-)protected group.
The fairness regularization, named as individual-based fairness gap (IFG) in \cite{Yan2019-iz}, is computed as: 
\begin{equation}
    \label{s3-eq:ifg}
    Loss_{IFG,t} =  \frac{1}{\sum_{i \in N} z_{i,t}^*}
                    \left|
                        \frac{\sum_{i \in N} z_{i,t} w_i^+}{\sum_{i \in N} p_i w_i^+} - 
                        \frac{\sum_{i \in N} z_{i,t} w_i^-}{\sum_{i \in N} p_i w_i^-}
                    \right|,
\end{equation}  

where $p_i$ is the total population of $s_i$ and $w_i^+$ ($w_i^-$) is the percentage of population in the protected (non-protected) groups. In this study, protected groups refer to Black and African-American, Hispanic and Latino, and Asian population. Non-protected group refers to non-Hispanic and non-Latino White population. We denote this second baseline as IFG; and the under-reporting-aware model proposed is denoted as TC (predicting crime hotspots based on inferred \textit{true crimes}).

\subsection{Accuracy and Fairness Metrics}
We measure the accuracy of the proposed under-reporting-aware model and the baselines using the F1 score. Specifically, we calculate the average monthly F1 score across the five test months for each model and city for both property and violent crimes. 

The fairness evaluation of short-term crime prediction models can be framed within the field of algorithmic fairness, which is based on the notion of protected groups.
A protected group represents a population sub-group that has historically suffered from discrimination and therefore some form of (approximate) parity or non-discrimination regulation in the predictive algorithm is desired for these groups \cite{Chouldechova2018-jv}. 
Since the discussion around algorithmic fairness for crime prediction has mostly focused on race and ethnicity \cite{Lum2016-or, Brantingham2017-ub, Kleinberg2017-yl}, in this study we evaluate the fairness of the short-term crime prediction results obtained with the proposed model and baselines with respect to three protected (minority) groups: 
Black or African-American (BA), Hispanic or Latino (HL), and Asian (A); and one non-protected (non-minority) group comprised of non-Hispanic and non-Latino Whites (W), as defined by the American Community Survey \cite{us_acs}.

Fairness is a complex concept and there are different metrics measuring different aspects of fairness. Given the problem setting as a binary classification - positive crime prediction means a census tract is likely to be a crime hotspot in the next day - and our focus on analyzing fairness using protected and non-protected groups, we apply four fairness metrics commonly used in the literature \cite{Kleinberg2017-yl, Lum2016-or, Verma2018-kq}: statistical parity (SP); false positive error rate balance (FPR); false negative error rate balance (FNR); and predicted positive to ground truth positive ratio (the metric used in the study by Lum and Isaac \cite{Lum2016-or}, LI). Achieving fairness with respect to these metrics means that the metric value for the protected group should be equal or similar to the metric value for the non-protected group. For example, if $SP_{pg} = SP_{npg}$ then the prediction model is considered to be fair in terms of statistical parity, where $pg$ stands for the protected group and $npg$ for the non-protected group. Next, I describe each fairness metric in detail and present how I use these metrics to compute a measure of fairness across protected groups. The four fairness metrics used in this paper are:

\textbf{Statistical Parity. $SP$} measures the fraction of the population in a (non-)protected group associated to a positive prediction {\it i.e.,} to a census tract with crime occurring the next day. Fairness in terms of SP suggests that the percentage of the population for a given group associated to positive predictions (crime) should be independent of the group itself, regardless of the ground truth crime data. If the $SP_{pg}$ is larger than $SP_{npg}$, then the short-term crime prediction model is biased towards protected groups, who would have a higher probability of being associated to crime in next-day predictions than non-protected groups. 
    
The SP metric is computed per (non-)protected group $g$ as follows: $SP_{g}=\frac{TP+FN}{TP+FP+TN+FN}$ where $TP$ in my setting is defined as the total population of the (non-)protected group $g$ associated to census tracts that were correctly predicted with crime occurring the next day across the whole testing period (August to December 2020); $TN$ is defined as the total population of the (non-)protected group $g$ associated to census tracts that were correctly predicted as non-crime hotspots for the next day across the testing period; $FP$ represents the total population of the (non-)protected group $g$ associated to census tracts that were incorrectly associated to crime occurring the next day across the testing period; and $FN$ refers to the total population of the (non-)protected group $g$ associated to census tracts that were incorrectly predicted as not having crime across the testing period.
    
\textbf{False positive error rate balance. $FPR$} measures the fraction of population in a (non-)protected group that is incorrectly associated to a positive prediction {\it i.e.,} to a census tract with predicted crime occurring the next day, despite the ground truth saying the opposite (no crime). Fairness in terms of $FPR$ suggests that the percentage of errors in the positive prediction should be independent of the population groups. If the $FPR_{pg}$ is larger than $FPR_{npg}$, the short-term crime prediction model is biased towards incorrectly making larger errors in the prediction of protected groups being involved in crimes. The $FPR$ metric is computed for each (non-)protectd group $g$ as follows: $FPR_{g}=\frac{FP}{TN+FP}$ with $FP$ and $TN$ defined as explained for the $SP$ metric.
    
\textbf{False negative error rate balance. $FNR$} measures the fraction of population in a (non-)protected group that is
incorrectly associated to a negative prediction {\it i.e.,} to a census tract without crimes predicted for the next day, despite the fact that the ground truth points to the presence of crime in that tract. Fairness in terms of $FNR$ suggests that the percentage of errors in the negative prediction should be independent of the population groups. If the $FNR_{pg}$ is smaller than $FNR_{npg}$, the short-term crime prediction model is biased in incorrectly believing that the non-protected group is less likely to be involved in crimes. The $FNR$ metric is computed for each (non-)protected group $g$ as follows:
$FNR_{g}=\frac{FN}{TP+FN}$ with $FN$ and $TP$ as explained in the $SP$ metric.
    
\textbf{Lum and Isaac. $LI$} measures the ratio between (1) the total population of a (non-)protected group associated to predicted crime hotspots by the short-term prediction model and (2) the total population of the same (non-)protected group associated to ground truth crime hotspots {\i.e.,} population in census tracts where crime occurrences are predicted versus the population for whom those crime occurrences are ground truth. Fairness in terms of $LI$ suggests that the (non-)protected groups are represented in the model predictions proportionally to the ground truth crime dataset. If $LI_{pg}$ is larger than $LI_{npg}$, the protected groups would be over-represented in the predicted hotspots when compared to the non-protected group. The $LI$ metric is computed for each (non-)protected group $g$ as follows:
$LI_{g}=\frac{TP+FP}{TP+FN}$ with $TP$ and $FP$ as explained in the $SP$ metric.

\textbf{Quantifying Changes in Fairness.} To quantify the degree of unfairness ($D$) of the short-term crime predictions from the proposed under-reporting-aware model (TC) and the three baselines (UU, UUC(C) and IFG), we propose the following approach. We calculate for each predictive model and fairness metric described the ratio between each pair of protected and non-protected group metric and subtract 1. The closer $D$ is to zero, the lower the degree of unfairness associated to the prediction.
For SP, FPR and LI,
positive $D$ values point to higher degrees of unfairness for the protected groups, while negative $D$ values point to higher degrees of unfairness for the non-protected groups. For example, $D_{BA/W,FPR}=\frac{FPR_{BA}}{FPR_{W}} - 1$ represents the degree of unfairness in crime prediction in terms of $FPR$ for the protected group BA (Black and African-American) compared to the non-protected group W (non-Hispanic and non-Latino White).
If $D_{BA/W,FPR}>0$ this reflects a higher degree of unfairness for Black and African-Americans when compared to non-Hispanic, non-Latino Whites. 
For FNR, positive (negative) $D$ point to higher degrees of unfairness for not-protected (protected) groups.

To measure the changes in fairness brought about by the proposed under-reporting-aware model (TC), we compare its degree of unfairness in crime prediction with the unfairness of the baselines (UU and IFG); and we use a 5\% relative change in the degree of unfairness as a threshold to  
determine whether the TC approach improves fairness compared to the baselines. 
If the ratio between the degree of unfairness of TC and UU (denoted as TC/UU) 
or between the degree of unfairness of TC and IFG (denoted as TC/IFG) is smaller than 0.95,
then we claim the TC approach improves fairness for a specific metric and protected group. 
For example, 
if $\frac{D_{BA/W,FPR,TC}}{D_{BA/W,FPR,UU}}<0.95$, the TC approach improves fairness for Black and African-American community in terms of false positive rate.

\section{Results}

    

\begin{figure}
    \centering
    \includegraphics[width=.75\linewidth]{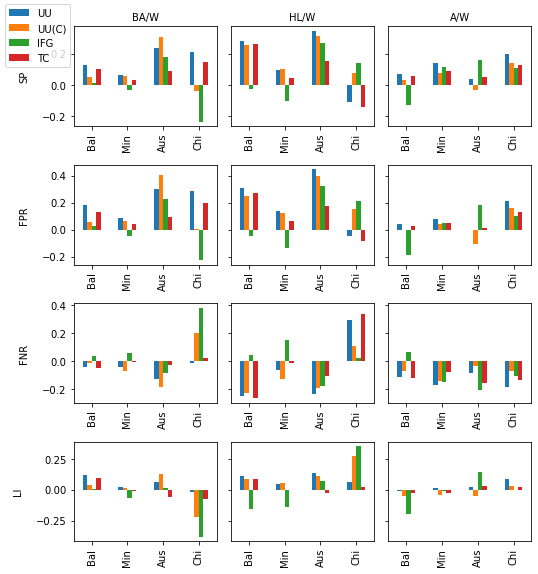}
    \caption{Degrees of unfairness of property crime prediction for four cities (Baltimore, Minneapolis, Austin and Chicago). Results are shown for each fairness metric explained in Section 5.3 and for each protected group. Crime prediction models include under-reporting-unaware model (UU), UU with historical crimes only (UU(C)), UU with individual-based fairness gap regularization, and the proposed under-reporting-aware model (TC).}
    \label{s3-fig:fair_tc_ppt}
\end{figure}

\begin{figure}
    \centering
    \includegraphics[width=.75\linewidth]{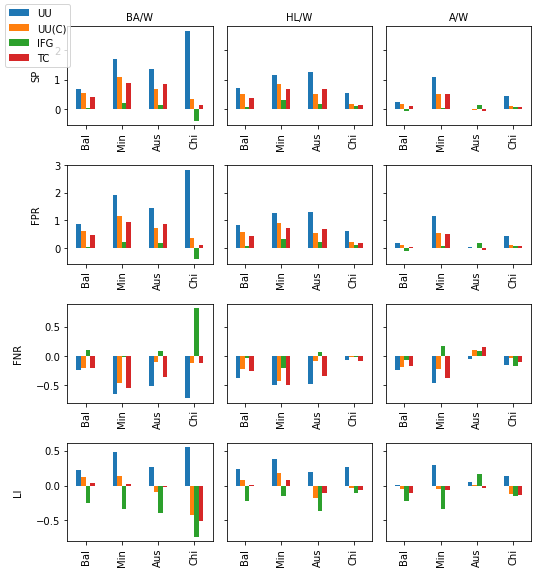}
    \caption{Degrees of unfairness of violent crime prediction for four cities (Baltimore, Minneapolis, Austin and Chicago). Results are shown for each fairness metric explained in Section 5.3 and for each protected group. Crime prediction models include under-reporting-unaware model (UU), UU with historical crimes only (UU(C)), UU with individual-based fairness gap regularization, and the proposed under-reporting-aware model (TC).}
    \label{s3-fig:fair_tc_vln}
\end{figure}

\subsection{Fairness}
Figure \ref{s3-fig:fair_tc_ppt} and \ref{s3-fig:fair_tc_vln} show the degree of unfairness of crime prediction for the four cities and the two types of crimes in terms of four fairness metrics for different race and ethnicity groups. They show that the effect on fairness resulting from adding a fairness regularization (IFG) or a convolutional gate for crime-reporting (TC) varies across cities and types of crimes. 
To be able to summarize the impact of fairness treatments, we calculate 
the percentage of settings for which the TC approach improves fairness compared to the baselines,
with a setting being defined as a combination of city, fairness metric and protected group. 
In other words, we compute the percentage of settings for which there is at least a 5\% relative change in the degree of unfairness of the crime prediction when using TC versus one of the other baselines. 
If the percentage of settings is larger than 50\%, applying the TC approach is beneficial in terms of improving fairness.
For example, our results show that when using the under-reporting-aware model (TC), 69\% of the times the fairness was improved when compared to an 
under-reporting-unaware model (UU) suggesting that among the cities, fairness metrics and race/ethnicity groups considered in this study, 69\% of the degrees of unfairness decrease more than 5\%.  
Next, we discuss the main findings.

\begin{table}[]
\centering
\begin{tabular}{lcccc}
\toprule
               & IFG/UU &  TC/UU & TC/IFG  & TC/UU(C) \\
\midrule
Property crime &  58\%   &  69\%  & 65\%   & 58\%    \\
Violent crime  &  71\%   &  85\%  & 40\%   & 54\%   \\
\bottomrule
\end{tabular}
\caption{Percentage of settings where applying the convolutional gate for crime-reporting process (TC) improves the baselines. }
\label{s3-tab:chance-fair-overall}
\end{table}

The results are shown in Table \ref{s3-tab:chance-fair-overall}. 
For property crime prediction, the IFG and TC approach both are beneficial to improving fairness (the percentages for IFG/UU and for TC/UU are both larger than 50\%) and the proposed TC approach has a better chance of improving fairness than IFG when compared to the UU baseline (69\% versus 58\%). In addition, the overall percentage of TC/IFG is 65\% suggesting the TC approach can further improve fairness than the baseline IFG approach. 

For violent crime prediction, both IFG and TC also are beneficial to improving fairness and similar with property crime prediction, TC has a better chance to improve over baseline UU than IFG (85\% for TC/UU versus 71\% vs IFG/UU).
However, comparing the TC approach with the baseline IFG approach, the chance for improvement (TC/IFG) is 40\% suggesting that the scale of reducing degrees of unfairness brought by TC tends to be smaller than the IFG approach in violent crime prediction. 
Interestingly, as we will discuss in the next section, this result reflects a trade-off between accuracy and fairness since the IFG has a larger decrease in F1 score for violent crimes than the TC approach (see Table \ref{s3-tab:avgf1_tc}, \textit{e.g.}, the decrease can be as large as 60\% in violent crime prediction for Austin (Aus) with F1 score 0.090 using IFG vs. 0.151 using TC).

Comparing property and violent crimes prediction, the percentage of conditions where fairness improves is larger for violent crimes (85\% for TC/UU) than for property crimes prediction (69\% for TC/UU). 
Prior work has shown that data bias is more severe for violent crime incidents than for property crime incidents \cite{wu2023auditing}. Hence, this result suggests that the more severe the data bias is, the higher the percentage of improving fairness fairness using the convolutional gate for crime-reporting (TC).

Comparing with UU model with historical crimes only, \textit{i.e.}, UU(C), the percentage of fairness improvement is over 50\% for both types of crimes. This means that the proposed TC approach can still slightly improve fairness over UU(C).

\begin{table}[]
\centering
\begin{tabular}{cccc}
\toprule
                               &      & Property crime & Violent crime \\
\midrule
\multirow{4}{*}{Metric}        & SP   & 83\%           & 92\%          \\
                               & FPR  & 83\%           & 92\%          \\
                               & FNR  & 50\%           & 75\%          \\
                               & LI   & 58\%           & 83\%          \\
\midrule
\multirow{3}{*}{Race/Ethnicity}& BA/W & 75\%           & 100\%         \\
                               & HL/W & 75\%           & 88\%          \\
                               & A/W  & 56\%           & 69\%          \\
\midrule
\multirow{4}{*}{City} & Bal & 67\%           & 92\%          \\
                      & Min & 92\%           & 92\%          \\
                      & Aus & 58\%           & 75\%          \\
                      & Chi & 58\%           & 83\%         \\
\bottomrule
\end{tabular}
\caption{Percentage of settings for which applying the convolutional gate for crime-reporting process (TC) improves fairness when compared with the under-reporting-unaware model (UU) by fairness metrics, race/ethnicity groups and cities.}
\label{s3-tab:chance-fair-tc}
\end{table}

Finally, since TC is the model that improves fairness the most - albeit at the cost of reducing accuracy as we will show in the next section- we take an in-depth look into the fairness improvement brought about by the TC approach versus the UU baseline by disaggregating the results by fairness metrics, by race/ethnicity groups and by cities as shown in Table \ref{s3-tab:chance-fair-tc}. We highlight three main results. 

First, in terms of fairness metrics, adding the convolutional gate for crime-reporting (TC approach) to the under-reporting-unaware model (UU approach) is especially good for improving fairness in terms of statistical parity (SP) and false positive rate (FPR) \textit{e.g.}, the chance of improving fairness is larger than 80\% in terms of SP and FPR for both types of crimes. This means that by modeling the data bias generated by the under-reporting issue, the percentage of population in the protected groups being involved in the predicted crime hotspots is reduced ({\it i.e.,} the degree of unfairness of SP decreases) due to less false positive prediction.

Second, for race/ethnicity groups, the chance to improve fairness is better for the Black and African-American as well as the Hispanic and Latino groups than for the Asian group. Prior work has shown that bias in crime data for US cities appears to be higher for Black and African-American as well as Hispanic and Latino groups \cite{wu2023auditing}. Hence, this result
suggests that the TC approach mitigates the data bias in reported crimes better for the protected race/ethnicity groups with more severe data bias. 

Third, the per-city values show the trade-off between accuracy and fairness, that is, larger chances of improving fairness generally correspond to larger decreases in prediction accuracy. For example, the decrease in prediction accuracy for Austin and Chicago in property crime prediction is small (F1 score decreases less than 0.02, see Table \ref{s3-tab:avgf1_tc}); while the chance of improving fairness is the smallest for Austin and Chicago (58\%). On the other hand, although the F1 score for Minneapolis in property crime prediction decreases by almost 0.1, the chance to improve fairness for Minneapolis is as high as 92\%.

\begin{table}[]
\small
\centering
\begin{tabular}{ccccccccc}
\toprule
\multicolumn{1}{l}{} & \multicolumn{4}{c}{Property crime} & \multicolumn{4}{c}{Violent crime} \\
                     & Bal     & Min    & Aus    & Chi    & Bal    & Min    & Aus    & Chi    \\
\midrule
UU(C)                & 0.345   & 0.533  & 0.533  & 0.309  & 0.369  & 0.299  & 0.124  & 0.210 \\
UU                   & 0.405   & 0.543  & 0.576  & 0.362  & 0.410  & 0.331  & 0.201  & 0.286  \\
IFG                  & 0.403   & 0.538  & 0.575  & 0.315  & 0.372  & 0.213  & 0.090  & 0.199  \\
TC                   & 0.386   & 0.444  & 0.559  & 0.362  & 0.366  & 0.287  & 0.151  & 0.215  \\
\bottomrule
\end{tabular}
\caption{Average monthly F1 score for property and violent crime prediction from Aug. to Dec. 2020 for each city. UU(C) means UU model but with historical crimes only as input features.}
\label{s3-tab:avgf1_tc}
\end{table}

\subsection{Accuracy} 
Table \ref{s3-tab:avgf1_tc} shows the F1 scores for property and violent crime prediction in the testing phase. It can be observed that both the baseline fairness improvement method (\textit{IFG}) and the proposed under-reporting-aware model (\textit{TC}) tend to have lower F1 scores across cities for both types of crimes. Put together with the results discussed earlier whereby TC and IFG have higher fairness metrics, this reflects a trade-off between accuracy and fairness. 
For IFG, the trade-off is due to the added fairness regularization,
that is, instead of minimizing solely the MSE (the error between predicted and ground truth number of reported crimes), IFG balances between the MSE and the fairness regularization which leads to an increase in the error and a decrease in the accuracy.
While for TC, since only reported crime data is available, there is no ground truth to evaluate the actual accuracy of the predicted \textit{true} crime hotspots. The mismatch between the predicted true crime hotspots and the ground truth reported crime hotspots might be the cause of decrease in accuracy. For example, a census tract with low reporting rate could be considered as a hotspot based on the inferred number of true crimes, but not a hotspot based on the reported crime data.
Finally, although TC has a decrease in accuracy compared with UU, TC still has similar accuracy with UU(C), which reflects that the improvement in accuracy by the additional mobility features has been offset by the convolutional gate mechanism.

\section{Conclusions}
Deep learning crime predictive tools use past crime data and additional behavioral datasets to forecast future crimes. Nevertheless, these tools have been shown to suffer from unfair predictions across minority racial and ethnic groups.  
Current approaches to address this unfairness generally propose either pre-processing methods that mitigate the bias in the training datasets by applying corrections to crime counts based on domain knowledge or in-processing methods that are implemented as fairness regularizers to optimize for both accuracy and fairness. 
In this paper, we have proposed a novel deep learning architecture that
combines the power of these two approaches to increase prediction fairness. 
Our results have shown that for property crime prediction, the proposed TC approach is beneficial to improving fairness, and has a better chance of improving fairness than the baseline with a simple fairness regularizer (IFG), when compared to the under-reporting-unaware baseline (UU). Our results also show that for violent crime prediction, the proposed method (TC) also improves fairness and, similarly to property crime prediction, TC has a better chance of improvement over the UU baseline than the IFG.
Interestingly, our results also reflect a clear trade-off between accuracy and fairness when comparing the accuracy of the under-reporting-aware and the baseline with the fairness regularizer against the UU baseline. 

\bibliography{ref} 

\end{document}